\documentclass[]{mn2e}
\pdfoutput=1
\usepackage{times}
\usepackage{epsfig}
\usepackage{amssymb}
\usepackage{amsmath} 
\def\lsim{\mathrel{\rlap{\lower4pt\hbox{\hskip1pt$\sim$}}
    \raise1pt\hbox{$<$}}}                
\def\gsim{\mathrel{\rlap{\lower4pt\hbox{\hskip1pt$\sim$}}
    \raise1pt\hbox{$>$}}}                
\title[The dust shell around SN\,2008S]
{The destruction and survival of dust in the shell around SN\,2008S}
\author[R. Wesson et al.]
{R. Wesson$^1$,  M. J. Barlow$^1$, 
B. Ercolano$^{1,2}$,
J. E. Andrews$^3$,
Geoffrey C. Clayton$^3$,
\newauthor
J. Fabbri$^1$,
Joseph S. Gallagher$^3$,
M. Meixner$^4$,
B. E. K. Sugerman$^5$, 
D. L. Welch$^6$,
\newauthor 
D. J. Stock$^1$\\
$^1$Department of Physics and Astronomy, University College London, Gower Street, London WC1E 6BT, UK\\
$^2$Institute of Astronomy, University of Cambridge, Madingley Road, Cambridge CB3 0HA, UK\\
$^3$Department of Physics and Astronomy, Louisiana State University, Baton Rouge, LA 70803\\
$^4$Space Telescope Science Institute,  3700 San Martin Drive, Baltimore, MD 21218\\
$^5$Department of Physics and Astronomy, Goucher College, 1021 Dulaney Valley Road, Baltimore, MD 21204\\
$^6$Department of Physics and Astronomy, McMaster University, Hamilton, Ontario L8S 4M1, Canada
}
\date{Received: 1st July 2009}

\begin{document}
\maketitle

\begin{abstract}

SN\,2008S erupted in early 2008 in the grand design spiral galaxy 
NGC~6946. The progenitor was detected by Prieto et al. in {\em Spitzer 
Space Telescope} images taken over the four years prior to the explosion, 
but was not detected in deep optical images, from which they inferred a 
self-obscured object with a mass of about 10M$_\odot$. We obtained {\em 
Spitzer} observations of SN~2008S five days after its discovery, as well 
as coordinated Gemini and {\em Spitzer} optical and infrared observations 
six months after its outburst.

We have constructed radiative transfer dust models for the object before 
and after the outburst, using the same r$^{-2}$ density distribution of 
pre-existing amorphous carbon grains for all epochs and taking 
light-travel time effects into account. 
We rule out silicate grains as a significant component of the dust around 
SN~2008S. The inner radius of the dust shell moved outwards from its 
pre-outburst value of 85~AU to a post-outburst value of 1250~AU, 
attributable to grain vaporisation by the light flash from SN~2008S. 
Although this caused the circumstellar extinction to decrease from A$_V$ = 
15 before the outburst to 0.8 after the outburst, we estimate that less 
than 2\% of the overall circumstellar dust mass was destroyed.

The total mass-loss rate from the progenitor star is estimated to have 
been 0.5-1.0$\times10^{-4}$~M$_{\odot}$~yr$^{-1}$.
The derived dust mass-loss rate of 5$\times10^{-7}$~M$_\odot$~yr$^{-1}$ 
implies a total dust injection into the ISM of up 
to 0.01~M$_\odot$ over the suggested duration of the self-obscured phase. 
We consider the potential contribution of objects like SN~2008S
to the dust enrichment of galaxies. 

\end{abstract}


\section{Introduction}

The unusual transient SN\,2008S was discovered on 1.8 February 2008 
(Arbour 2008) in the spiral galaxy NGC 6496, the ninth supernova to be 
discovered in that galaxy in the last hundred years.  Subsequently it was 
found by Botticella et al. (2009) to be present on an image taken on 24 
January 2008, but not on an image taken on 16 January 2008, allowing them 
to constrain the epoch of the explosion to (20$\pm$4) January 2008.

Prieto et al. (2008) reported that serendipitous pre-outburst deep optical 
observations of the supernova field did not reveal any source at its 
position down to 26th magnitude (Welch, Clayton \& Sugerman 2008 
provided further prediscovery deep optical photometric upper limits) and 
that archival {\em Spitzer Space
Telescope} IRAC images revealed a bright mid-infrared source whose 
brightness had not varied significantly over the four years preceding the 
outburst. The progenitor of SN\,2008S thus appeared to be enshrouded in a 
cloud of its own dust. From a 440~K blackbody fit to the IR fluxes, they 
estimated a pre-explosion luminosity of 3.5$\times10^4$~L$_\odot$ and a 
modest mass of $\sim$10~M$_\odot$ for the progenitor star, close to the 
poorly defined boundary between AGB and supergiant stars. They
suggested that SN~2008S was an electron capture SN event by such a star.
Wesson et al. (2008) reported IRAC and MIPS {\em Spitzer} 
observations of SN~2008S that were serendipitously obtained only 5 
days after its discovery, fitting the photometry with a 500\,K modified
blackbody with a luminosity of 2.1$\times10^6$~L$_\odot$. Although faint 
for a supernova, the amplitude and luminosity of the outburst of SN~2008S 
did not obviously correspond to any other class of object. 
Smith et al. (2009) presented optical photometry 
and spectroscopy of SN~2008S up to 280 days after outburst and
argued that it was a `supernova imposter' originating from
a super-Eddington outburst of a $\leq$15~M$_{\odot}$ star. Botticella 
et al. (2009) have presented a very large optical and infrared set of 
observations of SN~2008S up to day 304, and a model of the post-outburst light 
echo.  They argued that the late-time decline rate observed for SN~2008S was 
consistent with the decay of a relatively small quantity of $^{56}$Co produced 
by an electron capture supernova event in a super-AGB star of initial mass 
6-8~M$_\odot$. As noted by Thompson et al. (2008), Smith et al. (2009) and 
Botticella et al. (2009), if a supernova event has not occurred, a 
relatively luminous star should be detectable once the 2008 outburst has 
faded sufficiently.

The discovery that SN\,2008S had a dust-enshrouded progenitor led Thompson 
et al. (2008) to propose that it and two other recently discovered 
extragalactic transient sources were members of a new class of transient, 
formed when stars located near the upper mass limit for AGB evolution go 
through a self-obscured phase shortly before exploding as electron capture 
SNe.  Depending on the amount of dust produced in this phase, and the 
proportion of stars for which the phase occurs, objects such as 
SN~2008S could potentially contribute to the substantial 
amounts of dust observed in young high-redshift galaxies (e.g. Bertoldi et 
al. 2003), given that dust production by relatively massive stars 
($\gsim$5~M$_{\odot}$) appears to be required in order to match the short 
timescales over which dust is inferred to have formed in the early 
Universe (Morgan \& Edmunds 2003; Dwek, Galliano \& Jones 2009)

The utility of light echoes from circumstellar dust as a probe into the 
prior evolution of supernova progenitors has been discussed and exploited 
by e.g. Bode \& Evans (1980), Graham et al. (1983), Dwek (1983), Dwek 
(1985) and Sugerman (2003). The post-outburst evolution of the SED of
SN~2008S thus provides important information on the properties
of its progenitor. We present the mid-infrared observations of SN\,2008S 
that we obtained five days after its discovery, along with the further 
optical and infrared observations that we obtained in July 2008. We use 
these data to develop a self-consistent model for the evolution of the 
dust shell around SN\,2008S over three epochs, one pre-outburst and two 
post-outburst.

\section{Observations}

NGC 6946 was observed by {\em Spitzer} for the SINGS legacy survey, and 
also by programs monitoring the recent Type-II supernovae SN\,2002hh and 
SN\,2004et.  The field containing the progenitor of SN\,2008S was observed 
eight times in the four years preceding its eruption.  Prieto et al. 
(2008) reported the detection of the progenitor at 4.5, 5.8 and 8.0~$\mu$m.

While monitoring SN\,2002hh, we serendipitously obtained observations of 
SN\,2008S with IRAC and MIPS on 6.80 and 7.53 February 2008 respectively
(Wesson et al. 2008).
We subsequently reobserved the field with {\em Spitzer} in July 2008.  The 
object was observed in all four IRAC bands on 18 July, and with MIPS at 
24~$\mu$m on 29 July.  We measured fluxes from these images by fitting a 
synthetic psf to the observed profile, using the {\it daophot} package in 
{\sc iraf}. For the MIPS 24-$\mu$m data, we carried out psf fitting
on difference images that were produced by subtracting pre-outburst images 
from each of the two post-outburst images. We obtained an IRS spectrum of 
SN~2008S on 8 July 2008. It shows a featureless spectrum, lacking
obvious spectral lines or dust features. While of lower signal to
noise, its continuum level is consistent with the mid-IR  
photometry obtained during the same month.

We also observed SN\,2008S with the 8.1m Gemini North Telescope in Hawaii 
during July 2008.  We obtained 11.2-$\mu$m photometry with Michelle on 
7 July and optical photometry and spectra with GMOS on 14 July.  For the 
Michelle observations, the total on-source exposure time was 2400s, and 
the data were flux-calibrated using observations of the standard stars 
HD192781 and HD198149.

GMOS images of SN~2008S were taken in the g$'$, r$'$ and i$'$ filters, 
with an exposure time in each filter of 93\,s.  Johnson magnitudes of 
V$=21.25\pm0.05$, R$=20.14\pm0.04$ and I$=19.27\pm0.05$ were derived from 
the GMOS filter observations using the procedure described by Welch et al. 
(2007).

GMOS spectra were taken using the B600 grating, giving a spectral 
resolution of 1.8~{\AA} over a wavelength range of 4490-7356~{\AA}.  We 
obtained three spectra, dithering the central wavelength by 5\,nm between 
exposures to avoid gaps in the spectral coverage caused by gaps in the 
GMOS detectors.  The total exposure time was 3600s, and the spectrum was 
flux-calibrated using observations of the standard star BD+28 4211 (Massey 
et al. 1988).  The shape of the spectrum was in good agreement with the 
optical photometry, but since all spectroscopic observations were taken 
with a narrow slit, only relative flux calibration was possible, so we 
scaled our spectrum up to match the optical photometry.  Galactic 
foreground reddening in the direction of NGC 6946 is estimated to be 
E(B-V)=0.34 (Schlegel et al. 1998), and so we dereddened our optical 
photometry and spectrophotometry using this value.

Our July 2008 observations took place over a relatively narrow time 
interval, 169 days (Michelle), 176 days (GMOS), 180 days (IRAC) and 191 
days (MIPS) after the outburst of SN\,2008S, taken to be on 20 January 
(Botticella et al. 2009).  We combined our optical and mid-IR observations 
from February and July 2008 with the day 16/17 BVR photometry of Smith et 
al. (2009) and the day 174 JHK photometry of Botticella et al. (2009).

\begin{figure*}
 \epsfig{file=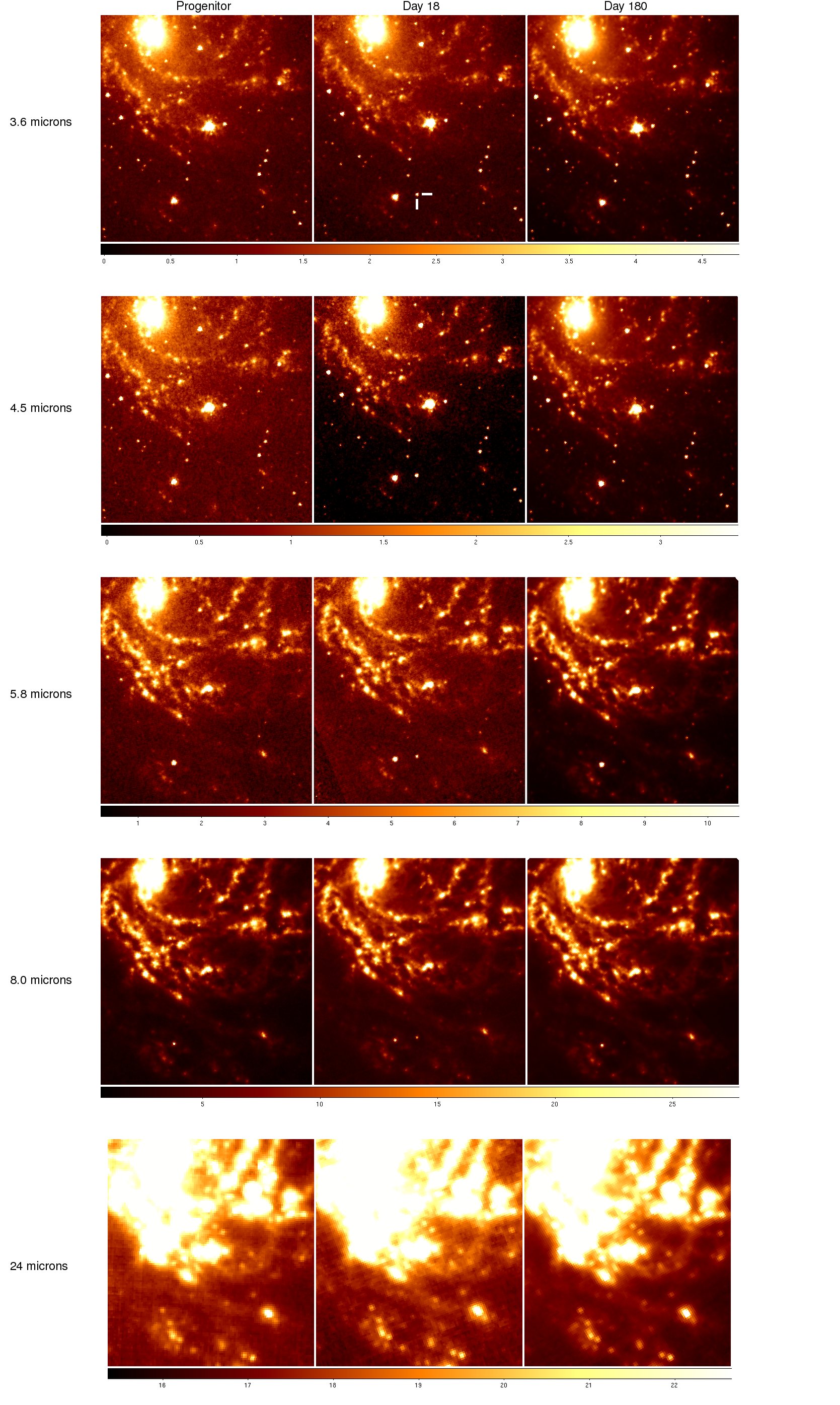, width=14cm}
 \caption{{\em Spitzer} images of the field around SN\,2008S in the four 
IRAC bands and in the MIPS 24-$\mu$m band, (a) before the 
eruption (left), (b) 17 days after after the eruption (centre), and (c) 
$\sim$180 days after the eruption, in July 2008 (right). The location of 
SN~2008S is indicated on the centre-top image.}
 \label{Spitzer_montage}
\end{figure*}

We present our measured fluxes from February 2008 and July 2008 in 
Table~\ref{Day17fluxes}.  Figure~\ref{Spitzer_montage} shows {\em Spitzer} 
pre-outburst and post-outburst images of SN\,2008S at five wavelengths 
from 3.6 to 24~$\mu$m.

\begin{table*}
\centering
\caption{Photometric measurements of SN\,2008S before, 17 days after, and 
six months after the explosion}
\label{Day17fluxes}
\begin{tabular}{ccccc}
\hline
Instrument & Wavelength ($\mu$m) & Progenitor flux (mJy)$^1$ & \multicolumn{2}{c}{SN 2008S flux (mJy)} \\
           &                     &                      & Day 17      & July 2008$^2$ \\
\hline
GMOS       & 0.55		 & -               & -           & 0.012$\pm0.01$ \\
           & 0.70                & -               & -           & 0.026$\pm0.01$ \\
           & 0.90                & -               & -           & 0.052$\pm0.02$ \\
IRAC       & 3.6                 & $<$0.0054       & 1.7$\pm$0.3 & 0.26$\pm$0.09 \\
           & 4.5                 & 0.022$\pm$0.004 & 2.0$\pm$0.3 & 0.42$\pm$0.12 \\
           & 5.8                 & 0.049$\pm$0.012 & 2.7$\pm$0.4 & 0.45$\pm$0.13 \\
           & 8.0                 & 0.066$\pm$0.013 & 4.0$\pm$0.4 & 0.83$\pm$0.18 \\
Michelle   & 11.2                & -               & -           & 1.0$\pm$0.5 \\
MIPS       & 24.0                & $<$0.096        & 0.85$\pm$0.25 & 0.56$\pm$0.20 \\
\end{tabular}
\\
{\small $^1$From Prieto et al. (2008). $^2$See Section~2 for dates of July 2008 observations.}
\end{table*}

\section{Modelling the spectral energy distributions}

\subsection{The Progenitor of SN~2008S}

Prieto et al (2008) fitted a black body curve to the mid-infrared spectral 
energy distribution (SED) of the progenitor, and found that it was fairly 
well fitted by a 440~K black body, giving an emitting radius of 150 AU and 
a luminosity of 3.5$\times$10$^4$\,L$_\odot$ for their adopted distance of 
5.6~Mpc.  In common with our studies of SN~2002hh (Barlow et al. 2005; 
Welch et al. 2007), we adopt a distance to NGC~6946 of 5.9~Mpc 
(Karachentsev et al. (2000), which if all other parameters were unchanged 
would yield luminosities that are 11\% higher than obtained by Prieto et 
al. and 7\% higher than obtained by Botticella et al. (2009). For SED 
fitting, it may be more realistic to assume a dust emissivity proportional 
to $\lambda^{-\alpha}$ in the mid-infrared, with $\alpha$ typically 
between 1 and 2.  For an $\alpha$\,=\,1 emissivity law, the progenitor's 
fluxes of Prieto et al. can be well fitted by a modified black body with a 
temperature of 380~K and a luminosity of 3.2$\times$10$^4$\,L$_\odot$ 
(Table~2).

\begin{table*}
\centering
\caption{Fits to the observed SN~2008S SEDs 
(corrected for a foreground Galactic extinction of A$_V$ = 1.05), 
at different epochs. Optical points were fitted by normal blackbodies, 
while mid-IR points were fitted by blackbodies modified by dust 
emissivities proportional to $\lambda^{-1}$. 
A distance to SN~2008S of 5.9~Mpc was adopted.}
\label{bbfits}
\begin{tabular}{cccccccl}
\hline
 & \multicolumn{3}{c}{Optical} & \multicolumn{2}{c}{Infrared} & Total & Ref\\
Epoch & T (K) & R (R$_\odot$) & L (L$_\odot$) & T (K) & L (L$_\odot$) & L (L$_\odot$) & \\
\hline
Progenitor & -    & -    & -                 & 440 (unmodified) & 3.6$\times$10$^4$ & 3.6$\times$10$^4$ & Prieto et al (2008)\\
           & -    & -    & -                 & 380 & 3.2$\times$10$^4$ & 3.2$\times$10$^4$ & This work\\
Day 17     & 6000 & 3360 & 12.9$\times$10$^6$& 550 & 2.2$\times$10$^6$ & 1.5$\times$10$^7$ & This work\\
Day 180$^1$  & 4000 & 1900 & 8.2$\times$10$^5$ & 450 & 3.6$\times$10$^5$ & 1.2$\times$10$^6$ & This work\\
\hline
\end{tabular}
\\
{\small $^1$July 2008; see Section~2 for dates of observations.}
\end{table*}

We used this luminosity as a starting point from which to construct 
increasingly detailed models of the dust shell surrounding SN\,2008S, 
experimenting with both amorphous carbon and silicate grains, as well
as mixtures of the two.  We 
constructed models of SN\,2008S and its circumstellar dust using the 
three-dimensional radiative transfer code {\sc mocassin} (Ercolano et al. 
2003, 2005, 2008). We began with the simplest case of a homogeneous 
constant density shell, obtaining a match to the observed mid-IR fluxes by 
varying the inner and outer radii, R$_{\rm in}$ and R$_{\rm out}$, and the 
dust mass.  We assumed that the central source had a luminosity of 
3$\times$10$^4$\,L$_\odot$ and a temperature of 10000\,K.  The high visual 
optical depth of the dust shell being considered means that the resulting 
SED is insensitive to the choice of central star temperature. 
For the amorphous carbon models, we used the 
grain optical constants of Hanner (1988), with particle 
sizes following a standard MRN distribution (Mathis, Rumpl \& Nordsieck 
1977) with a$_{min}$=0.005$\mu$m and a$_{max}$=0.25$\mu$m; the best match 
to the observations was obtained with R$_{\rm in}$ = 75~AU and R$_{\rm 
out}$ = 425~AU, and a dust mass of 2$\times$10$^{-5}$\,M$_\odot$. These 
inner and outer radii are similar to those estimated by Botticella et al. 
(2009) for the progenitor's dust shell.

We then constructed models in which the dust density in the shell was 
inversely proportional to the square of the radius.  This would be 
expected if the dust was forming continuously in an outflow from the star.  
We varied the inner and outer radii, and the total dust mass in the shell, 
to obtain a good fit.  We found that a standard MRN dust size distribution 
with grain sizes between 0.005 and 0.25\,$\mu$m could not reproduce the 
steepness of the observed SED between 3.6 and 4.5\,$\mu$m.  Prieto et al. 
(2008) suggested that the shape of the SED required the presence of large 
dust grains, and we found an improved fit by using an MRN distribution 
with grain sizes between 0.1 and 0.5\,$\mu$m.

The match to our observations is quite sensitive to the adopted inner 
radius, which is found to be 85~AU.  Unfortunately, the constraints on the 
outer radius of the dust shell are weak, as this only significantly 
affects the SED at wavelengths of $\sim$10$\mu$m and longer.  We therefore 
considered two alternative physically plausible models.  In the first 
case, we considered a dust shell with an outer radius 5 times the inner 
radius.  For an expansion velocity of $\geq$70~km~s$^{-1}$ (consistent 
with the late-time [Ca~{\sc ii}] line widths; see Section 
3.4), such an outer radius would correspond to an age of less than 30 
years.

\begin{figure}
 \epsfig{file=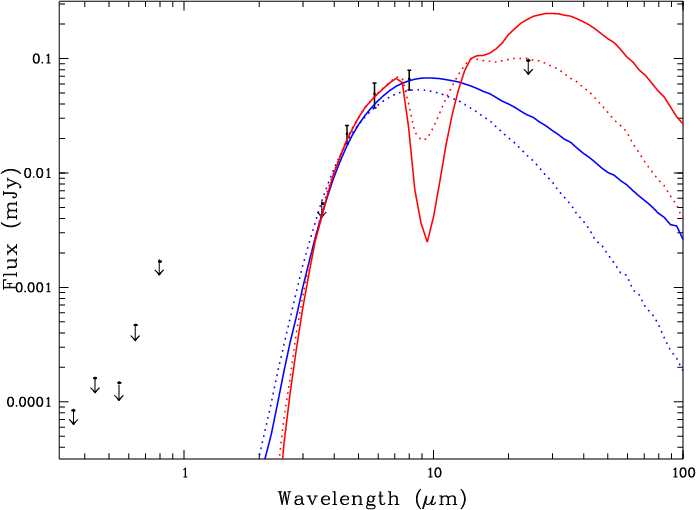, width=8cm}
 \caption{Fits to the SN~2008S progenitor SED using r$^{-2}$ dust density
distributions (see Table~3).  The optical and infrared
observations are from Prieto et al. (2008) and have been corrected for 
foreground Galactic extinction. The blue lines
are amorphous carbon (AC) models; the red lines are silicate
models.  Solid lines represent shells extending from 85-100,000\,AU
(AC), or from 85-20,000\,AU (silicates), while the dotted lines are 
for shells extending from 85-425\,AU.}
 \label{progenitor_models}
\end{figure}

Thompson et al. (2008) estimated that the dust-enshrouded phase for a star 
similar to the progenitor of SN~2008S could last up to ten thousand years.  
We therefore constructed alternative models of the dust shell with the 
same inner radius but with outer radii of 20,000~AU and 100,000~AU, 
corresponding respectively to the distances from the star reached
by a 70\,km\,s$^{-1}$ outflow after 1,500 years and 7,500 years. 
All of these models gave a good fit to the 
observed SED, and lie below the upper limit to the 24-$\mu$m flux.  The 
total dust mass is 1.2$\times$10$^{-5}$\,M$_\odot$ for R$_{\rm out}$ = 
425~AU, and 3.49$\times$10$^{-3}$\,M$_\odot$ for R$_{\rm out}$ = 100,000~AU.  
Henceforth we refer to the model with R$_{\rm out}$ = 425~AU as the 
`small' dust shell, and the R$_{\rm out}$ = 100,000~AU model as the `large' 
dust shell. Figure~\ref{progenitor_models} compares the SEDs predicted by 
these two models with the Prieto et al. (2008) optical and infrared 
observations of the progenitor.

We also investigated whether silicate dust shells could account for the 
observed SED of the progenitor of SN~2008S.  Using the `astronomical 
silicate' optical constants of Draine \& Lee (1984), the dust mass 
required to approximately match the SED was considerably higher than for 
amorphous carbon grains, due to the greater transparency in the visible 
and near-IR of silicate grains.  A full match to the observations was 
impossible because the dust masses required to match the steepness of the 
SED between 3.6 and 4.5~$\mu$m give rise to a strong, broad silicate 
absorption feature at 10~$\mu$m, which is inconsistent with the observed 
8$\mu$m flux.  Silicate dust shells with larger outer radii of 
20,000\,AU or 100,000~AU also predicted a 24-$\mu$m flux much higher than 
the observed upper limit for the progenitor of SN\,2008S 
(Figure~\ref{progenitor_models}).

With silicaceous dust thus ruled out for the progenitor's SED, we modelled 
the dust shell at subsequent epochs using carbon dust having the same 
density distribution and the same minimum and maximum grain radii as the 
progenitor `large' dust shell model. The progenitor model input parameters 
are summarised in Table~\ref{model_parameters} for the large and small 
shell models, for both silicaceous and carbon dust. Although we 
constructed models using both amorphous carbon and graphite, using 
graphite optical constants from Draine \& Lee (1984), all dust masses 
quoted are for the case of amorphous carbon; graphite grains required dust 
masses 7\% higher in each case.

\begin{table*}
\centering
\caption{Input parameters for r$^{-2}$ density distribution dust models  
for SN\,2008S, before, 17 days after, and six months after the explosion.  
In all cases the grain size a is proportional to a$^{-3.5}$, with a$_{max}$=0.5$\mu$m 
and a$_{min}$=0.1$\mu$m.}
\label{model_parameters}
\begin{tabular}{llllllllllll}
\hline
Epoch  & \multicolumn{8}{c}{Dust distribution}  & \multicolumn{2}{c}{Central source} \\
       & \multicolumn{2}{c}{R$_{in}$} & \multicolumn{2}{c}{R$_{out}$} & $\tau_V$ & Total mass (M$_\odot$) & Visible fraction & Composition & L(L$_\odot$) & T(K) & R(R$_\odot$) \\
       & (AU) & (light days) & (AU) & (light days) \\
\hline
Progenitor & 85 & 0.49   & 425 & 2.45      & 11.2 & 1.2$\times$10$^{-5}$  & 1.00 & AC$^1$ & 2.6$\times$10$^4$ & 10\,000 & 53.7\\
           & 85 & 0.49   & 100\,000 & 577.6& 13.5 & 3.49$\times$10$^{-3}$ & 1.00 & AC$^1$ & 3.3$\times$10$^4$ & 10\,000 & 60.5 \\
           & 85 & 0.49   & 425 & 2.45      & 188.0& 3.3$\times$10$^{-4}$  & 1.00 & Silicates$^2$  & 3.9$\times$10$^5$ & 10\,000 & 208 \\
           & 85 & 0.49   & 20\,000  & 115.5& 225.4& 1.9$\times$10$^{-2}$  & 1.00 & Silicates$^2$  & 5.2$\times$10$^5$ & 10\,000 & 240 \\
\\
Day 17     & 1250 & 7.22 & 100\,000 & 577.6& 0.76 & 3.45$\times$10$^{-3}$ & 0.07 & AC$^1$ & 1.57$\times$10$^7$ & 8\,000 & 2060 \\
\\
Day 180    & 1250 & 7.22 & 100\,000 & 577.6& 0.76 & 3.45$\times$10$^{-3}$ & 0.45 & AC$^1$ & \multicolumn{3}{c}{varying; see text} \\
\hline
\end{tabular}
\\
{\small $^1$Amorphous carbon; optical constants from Hanner (1988). $^2$Astronomical silicates; optical constants from Draine \& Lee (1984).}
\end{table*}

\subsection{SN~2008S at Day 17}

\begin{figure}
 \epsfig{file=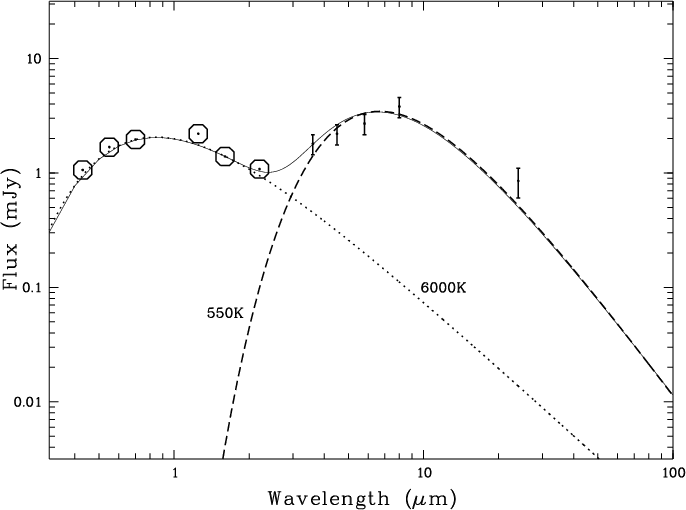, width=8cm}
 \caption{A 6000~K blackbody plus 550~K modified blackbody fit to the 
day 17 SED (corrected for foreground Galactic extinction) of 
SN\,2008S. The {\em Spitzer} photometry is from Table~1, the optical 
photometric points are from Smith et al. (2009) and the near-IR 
photometric points were extrapolated from day 41 by Botticella et al. 
(2009).}
 \label{day17_bbs}
\end{figure}

Our {\it Spitzer} observations of SN\,2008S took place on 6.8 February 
2008 (IRAC) and 7.5 February (MIPS), 17$\pm$4 days after the explosion.  
We combined our mid-IR flux measurements with the optical observations 
reported by Smith et al. (2009) from 6.1 February (B- and V-band) and
6.7 February
(R-band). We also followed Botticella et al. (2009) in extrapolating their 
day 41 JHK photometry back to day 17. We dereddened the photometry using 
the estimated Galactic foreground reddening of E(B-V)=0.34 (Section~2). 
The resulting SED can be well fitted by the sum of a 6000\,K blackbody 
plus a 550\,K blackbody modified by a $\lambda^{-1}$ emissivity law 
(Figure~\ref{day17_bbs}).  The luminosities of the two components are 
12.9$\times$10$^6$\,L$_\odot$ and 2.2$\times$10$^6$\,L$_\odot$ 
respectively, giving a total luminosity of 15.1$\times$10$^6$\,L$_\odot$.

Smith et al. (2009) estimated that the peak luminosity of SN~2008S was 
3$\times$10$^7$\,L$_\odot$. The flash would easily have vaporised dust at 
the inner edge of the shell.  The removal of the inner edge of the shell 
would then expose more dust to the intense radiation field, and dust 
destruction could proceed rapidly outwards.  To investigate how much of 
the shell could have been destroyed in this way, we ran a series of models 
with successively larger inner radii.

A sublimation temperature of 2200\,K is often assumed for amorphous 
carbon.  We found that for this value, the dust shell surrounding 
SN\,2008S could have been hollowed out to a radius of 215\,AU.  This 
resulted in a new optical depth of 2.1 for the small shell and 4.5 for 
the large shell.  Using this hollowed shell and varying only the source 
parameters, it was not possible to match the day 17 observations, 
as the optical depth was too high and gave too much reprocessed radiation.  
We therefore investigated what new inner radius was required to match the 
day 17 observations, and what dust sublimation temperature would be 
required to allow the shell to be hollowed out to this radius.

We found that a radius of 1250~AU was required to give dust cool enough at 
the inner edge, and a low enough optical depth to match the observations.  
The mid-infrared observations 
on day 17 and in July 2008 require that there still be a significant 
amount of dust remaining around the supernova.  Therefore, we can rule out 
the progenitor `small' shell model, as this would have been completely 
vaporised by the explosion, with no resulting mid-IR excess post-outburst.

For the Smith et al. (2009) peak luminosity of 3$\times$10$^7$\,L$_\odot$, 
the required dust sublimation temperature is 1050~K if dust inside 1250~AU 
is to be vaporised. Experimentally, the sublimation temperature of solid 
carbon varies from 2870~K for a carbon partial pressure of 1~Pa 
(10$^{-5}$~bar) to 1850~K at 10$^{-7}$~Pa (Grigoriev \& Meilikhov 1997).
An extrapolation of their relation to 10$^{-11}$~Pa, corresponding to the
gas density of $\sim$10$^5$~cm$^{-3}$ at 1250~AU (Section 4) combined with 
a carbon species abundance of 10$^{-3}$, predicts a carbon sublimation 
temperature of 1440~K. For this value of the grain sublimation temperature
a peak outburst luminosity of 1.5$\times10^8$~L$_\odot$ would 
have been required in order to hollow out the dust shell to 1250~AU.

For a shell in which the dust density is proportional to r$^{-2}$, each 
annulus of constant thickness contains the same mass.  Thus, hollowing out 
the shell to a new inner radius of 1250~AU only removes a small fraction 
of the original mass.  However, the optical depth is determined by the 
column density of the dust, proportional to R$_{\rm in}^{-1}$ 
- R$_{\rm out}^{-1}$ for an 
r$^{-2}$ shell, and so a dust shell hollowed out by a radiation flash can 
have a much lower optical depth but a similar mass to the pre-existing 
shell.  The mass of the `large' shell, originally 
3.49$\times$10$^{-3}$\,M$_{\odot}$, is reduced to 
3.45$\times$10$^{-3}$\,M$_{\odot}$ when hollowed out to 1250~AU.  
Meanwhile, its visual optical depth is reduced from 13.5 to 0.76.

\begin{figure}
 \epsfig{file=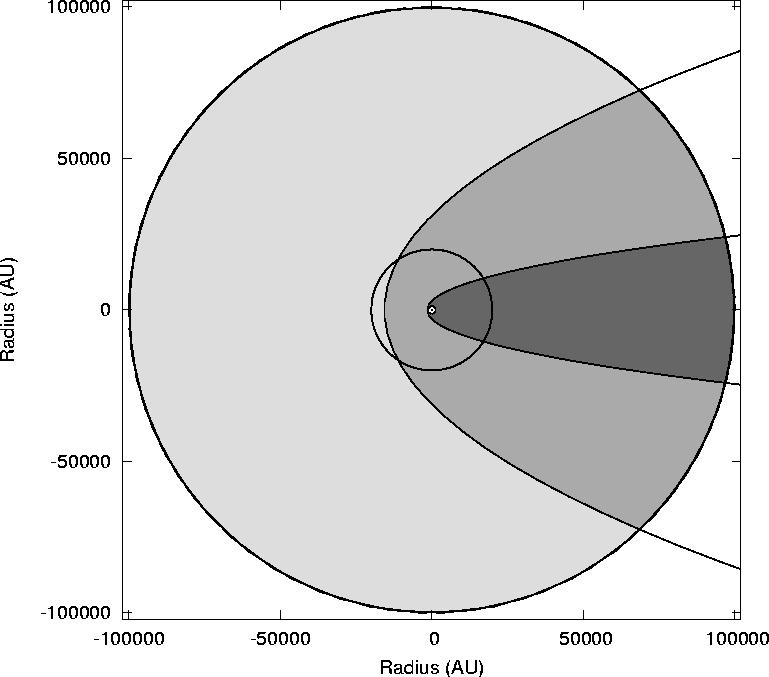, width=8cm} 
 \caption{Schematic of the dust shell around SN~2008S, 17 days and 180 days after 
outburst.  The darker shaded regions correspond to the parts of the dust shell from 
which reprocessed radiation has reached the observer, 17 days (inner ellipsoid) and 
180 days (outer ellipsoid) after the outburst. Dust inside the 
central cavity, of radius 1250~AU, has been vaporised by the initial 
light flash from the outburst. This schematic shows shells with outer radii
of 20,000~AU and 100,000~AU; our final model had an outer radius of 100,000~AU.}
 \label{lighttraveltime}
\end{figure}

A consideration of light travel time is crucial to modelling the observed
SED on day 17.  From the point of view of the observer, only dust lying
within an ellipsoidal region defined such that the
supernova-ellipsoid-observer distance is 17 light days greater than the
SN-observer distance will have been heated by the flash from the supernova
(Figure~\ref{lighttraveltime}). Emission reaching the observer from dust 
outside this ellipsoid left the dust before the supernova flash reached 
it.  The mass illuminated by the flash after 17 days is less than 7 per 
cent of the total mass of the shell, for an outer radius of 100,000~AU.  

To account for this effect, we created a customised version of MOCASSIN in
which the SED was constructed by summing emission only from those cells
lying within the ellipsoid described above, viewed from the direction in
which the ellipse opens out.  We assumed a constant luminosity
and temperature for the illuminating source over the first 17 days.  The
bolometric luminosity estimates of Botticella et al. (2009) indicate that
the source luminosity changed by a factor of $\le$1.25 over this period.

\begin{figure}
 \epsfig{file=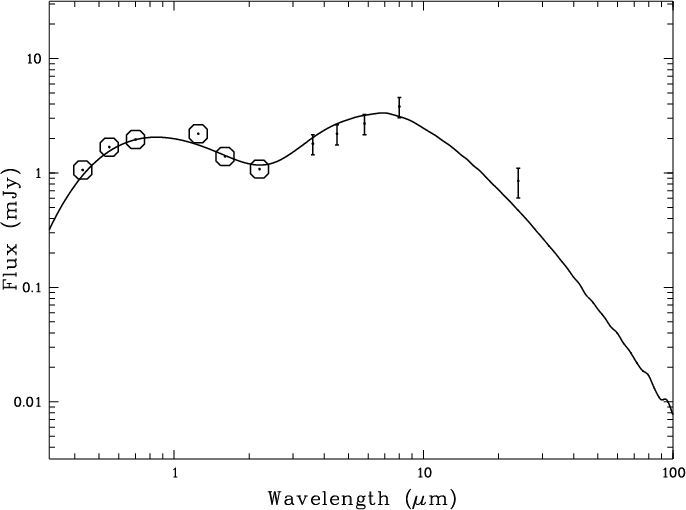, width=8cm}
 \caption{Observed fluxes (corrected for foreground Galactic 
extinction) and model predictions, 17 days after the 
outburst of SN~2008S. The photometric points are the same as for Figure~3.
The amorphous carbon dust shell model is the same as for the progenitor 
`large' shell model but with an inner radius of 1250~AU and powered by a 
8000~K central source with a luminosity of 1.57$\times10^7$~L$_{\odot}$.}
 \label{day17_model}
\end{figure}

With an inner radius of 1250~AU, we obtained a fit to the observed day~17 
SED using an 8000\,K source with a luminosity of 
15.7$\times$10$^6$\,L$_\odot$ (Table~2), as shown in 
Figure~\ref{day17_model}; details of the model are given in Table~3.
The fit can be considered reasonable; it lies somewhat below our 
measured  24-$\mu$m flux level for day~17, but this flux has a signal to 
noise ratio of only 3.4. For their day 17 SED modelling, Botticella et al. 
used single-radius (0.5~$\mu$m) carbon grains having an r$^{-2}$
density distribution between R$_{in}$ = 2000~AU and R$_{out}$ = 20000~AU.

\subsection{The July 2008 SED of SN~2008S}

The dereddened SED of SN\,2008S in July 2008, six months after outburst, 
can be fitted by the sum of a 4000\,K blackbody plus a 450\,K blackbody 
modified by a $\lambda^{-1}$ emissivity law, with luminosities of 
8.2$\times$10$^5$\,L$_\odot$ and 3.6$\times$10$^5$\,L$_\odot$, 
respectively.  The combined luminosity is then 
1.2$\times$10$^6$\,L$_\odot$ (Table~2), corresponding to a fading of a 
factor of 12.5 between day 17 and day 180.

We modelled the July 2008 fluxes using the dust distribution derived for 
the progenitor but again removing the dust in the inner regions, out to a 
radius of 1250~AU, which had been vaporised by the outburst according to 
our day 17 modelling.  By July 2008, 45 per cent of the 100,000AU shell has 
been illuminated by the outburst of SN~2008S.  To account for light travel 
time, we constructed a 
model consisting of five nested ellipsoids, each corresponding to an 
interval of 36 days, with the luminosity and temperature of the 
illuminating source declining from the outer to inner ellipsoids.  The 
input luminosities were derived by averaging the bolometric light curve 
presented in Figure~3 of Botticella et al. (2009) over the corresponding 36 
day period.  The central source temperatures were taken from Table~8 of 
Botticella et al. (2009); however, the model output was found 
to be insensitive to the adopted temperature. The fraction of the 
shell seen to be illuminated by the flash after 180 days is 44.7 per cent.

The SED produced by this model gave a good fit to the observed optical 
and IR SED (see Table~3 and Figure~6). Botticella et al. (2009) invoked dust 
formation in order to account for the observed near-IR photometric flux 
levels after day~120, but we find that our outburst-heated progenitor wind 
model can account for the observed near-IR flux levels in July 2008 
($\sim$day~180).

\begin{figure}
 \epsfig{file=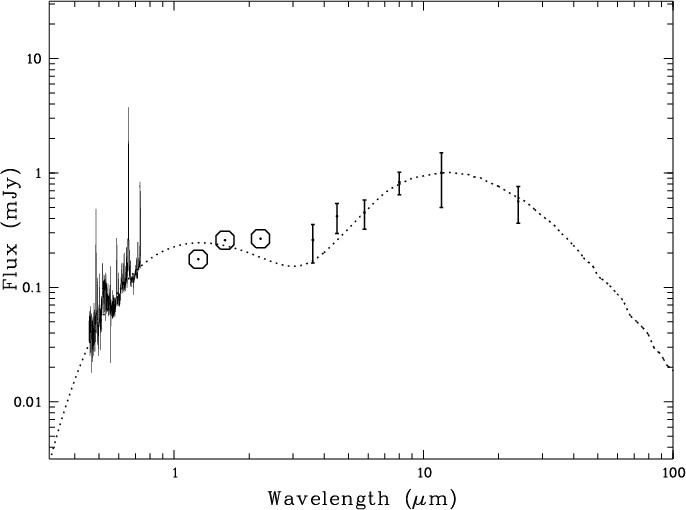, width=8cm}
 \caption{The observed and modelled SED of SN~2008S in July 2008. The
observations (corrected for foreground Galactic extinction) include our 
GMOS optical spectrophotometry, our {\em Spitzer} 
and Michelle mid-IR photometry (Table~1), and the day 174 JHK 
photometry of Botticella et al. (2009). The amorphous carbon dust shell 
model is the same as for the progenitor `large' shell model but with an 
inner radius of 1250~AU and powered by a 4000~K central source with a 
luminosity of 1.3$\times10^6$~L$_{\odot}$.
}
 \label{july08_models}
\end{figure}

\subsection{The July 2008 GMOS Spectrum}

\begin{figure*}
 \epsfig{file=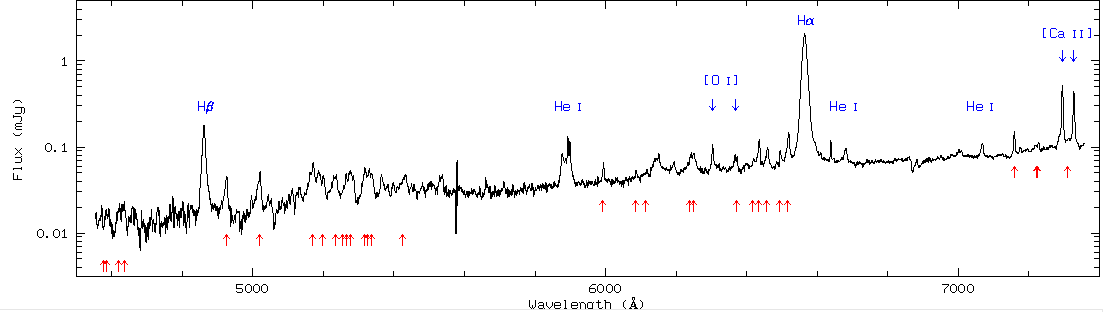, width=17cm}
 \caption{14 July 2008 GMOS-N optical spectrum of SN\,2008S, 175 days 
after outburst. Identifications for some of the stronger emission lines
are marked. The upward arrows indicate lines of Fe~{\sc ii}.}
 \label{spectrum}
\end{figure*}

Our 14 July 2008 (day 175) spectrum of SN~2008S is shown in 
Figure~\ref{spectrum}, with identifications for some of the stronger 
emission lines indicated. A comprehensive list of emission line 
identifications for SN~2008S can be found in Table~A2 of Botticella et al. 
(2009). As noted previously by Smith et al. (2009) and Botticella et al. 
(2009), the most striking aspect of the post-outburst spectra of SN~2008S 
is the narrowness of the H$\alpha$ line: we measured its full width at 
half-maximum (FWHM) to be 690\,km\,s$^{-1}$, much lower than the line 
widths of several thousand km\,s$^{-1}$ typically observed after the 
outbursts of CCSNe. The likely origin for the observed emission lines 
is the gaseous component of the dusty circumstellar outflow that 
predated the SN~2008S outburst. Botticella et al. (2009) observed 
smooth broad wings ($\sim$3000~km~s$^{-1}$ FWHM) in several lines, 
including H$\alpha$, in their early post-outburst spectra, which may have
been caused by electron scattering, although an underlying high-velocity 
outflow is a possibility.
Overall, they identified three line-width components, broad, intermediate 
and narrow, whose widths tended to decline with time. We could obtain a 
good fit to our day~175 H$\alpha$ profile with three such components: a 
broad component whose FWHM was 1460~km~s$^{-1}$, accounting for 25.1\% of 
the total flux; an intermediate component of FWHM 640~km~s$^{-1}$, 
accounting for 71.2\% of the flux; and a narrow component of FWHM 
165~km~s$^{-1}$, accounting for the remaining 3.7\% of the flux. The 
[Ca~{\sc ii}] 7291,7324~\AA\ lines had a dominant `narrow' component of 
FHWM 140~km~s$^{-1}$ (all quoted FWHMs were corrected in quadrature for 
the instrumental resolution of 1.8~\AA\ FWHM).

If the intermediate and narrow emission line components originate in a 
spherically symmetric circumstellar outflow, then their half widths at 
half-maximum should be approximately equal to the expansion velocity in 
the region(s) where they originate. The day~175 H$\alpha$ intermediate 
component then yields an expansion velocity of 320~km~s$^{-1}$, while the 
narrow H$\alpha$ component and the forbidden [Ca~{\sc ii}] 7291, 7324~\AA\ 
lines, which may originate from lower density material, further out in the 
wind, yield expansion velocities of 70-80~km~s$^{-1}$. The 
intermediate-width velocity component may correspond to material in the 
innermost regions of the wind that has been accelerated by the 
radiation pressure pulse from the outburst.

\section{Discussion: SN~2008S-type sources as significant dust 
contributors?}

Our analysis of SN\,2008S shows that both the progenitor SED and the day 
17 and day 180 SEDs can be self-consistently fitted by the same r$^{-2}$ 
density distribution of amorphous carbon grains, whose pre-outburst inner 
shell radius of R$_{\rm in}$ = 85~AU increased to 1250~AU after the 
outburst, due to grain evaporation by the light flash.  As a result of 
this vaporisation of the innermost grains, the visual optical depth of the 
dust shell ($\propto$ R$_{\rm in}^{-1}$ - R$_{\rm out}^{-1}$) decreased 
from 13.5 to 0.76, but the total mass of dust ($\propto$ R$_{\rm out}$ - 
R$_{\rm in}$) decreased from its intial value by less than 2\%. For an 
expansion velocity of 70~km~s$^{-1}$, our derived parameters imply a dust 
mass-loss rate of 5.2$\times10^{-7}$~M$_{\odot}$~yr$^{-1}$ from the 
progenitor of SN~2008S. If the gas to dust mass ratio in the outflow is 
100-200, then this corresponds to an overall mass-loss rate of 
0.5-1.0$\times10^{-4}$~M$_{\odot}$~yr$^{-1}$. Matsuura et al. (2009) find 
a high fraction of carbon-rich stars amongst the most luminous AGB stars 
in the LMC, with over ten such stars having inferred mass-loss rates of 
0.5-1.0$\times10^{-4}$~M$_{\odot}$~yr$^{-1}$, similar to that inferred 
here for the progenitor of SN~2008S. The total mass of the dust shell 
around SN~2008S could be larger than the 3.5$\times$10$^{-3}$\,M$_\odot$ 
in our dust model of outer radius 100,000~AU, since the total mass is 
proportional to the outer radius, which is not constrained by the 
available data. For our derived dust density at 1250~AU of 
1.1$\times10^{-19}$~g~cm$^{-3}$, the gas particle density n$_{\rm gas}$ 
there is (0.5-1.0)$\times10^5$~cm$^{-3}$, implying that n$_{\rm gas}$ = 
(1-2)$\times10^9$~cm$^{-3}$ at 10~AU, the effective emitting radius of the 
central source on both day 17 and day 180, and (1-2)$\times10^7$~cm$^{-3}$ 
at 85~AU, the presumed dust condensation radius of the progenitor wind. 
The latter density is similar to that inferred by Cherchneff et al. (2000) 
for the region of dust formation in late-type carbon-rich Wolf-Rayet 
outflows.

While Botticella et al. (2009) adopted a model in which SN~2008S had 
undergone episodic mass loss and a sharp change in dust chemistry, fitting 
the pre-outburst SED with silicate grains and the post-outburst SED with
carbon grains, we find that a featureless grain type such as amorphous 
carbon is required to fit both the 
pre-outburst and post-outburst SEDs of SN~2008S. This 
appears surprising at first sight, given that outflows from cool 
supergiants or super-AGB stars might be expected to be oxygen-rich, thus 
favoring the formation of silicate grains. However, we find silicate 
models to be completely incompatible with the observed optical and IR 
SEDs, producing a strong 10-$\mu$m absorption feature for high-$\tau_V$ 
pre-outburst models and a strong 10-$\mu$m emission feature for 
lower-$\tau_V$ post-outburst models. Comparison of our models for the July 
2008 SED of 
SN~2008S with the {\em Spitzer} and Michelle mid-IR photometry and the {\em 
Spitzer} IRS spectrum constrain any silicate grain component to contribute 
significantly less than 10\% of the overall dust mass. Since cool M 
supergiant stars always show silicate dust features in the mid-IR, the 
absence of silicates in the outflow around SN~2008S and its progenitor may 
count as evidence against a supergiant precursor and in favor of a 
super-AGB star progenitor. Surface carbon enrichment of AGB stars via the 
3rd dredge-up can produce carbon stars. However, at the higher mass end of 
the AGB mass distribution, envelope burning is predicted to convert 
dredged-up carbon to nitrogen, lowering the C/O ratio below unity and 
allowing silicate grains to form. Given the absence of silicate grains 
around SN~2008S, a mechanism to suppress or bypass envelope burning in 
super-AGB stars may be necessary, perhaps related to the severe mass loss 
stripping of the envelope.

Prieto et al. (2009) have obtained a post-outburst {\em Spitzer} IRS 
spectrum of the 2008 optical transient in NGC~300 (Monard 2008), whose 
evolution has shown many similarities to that of SN~2008S (Bond et al. 
2009). Prieto et al. find two strong emission
features in its mid-IR spectrum, which they attribute to carbon-rich particles, 
from a comparison with the mid-IR spectra of a number of C-rich post-AGB
objects. As well as this similarity between the inferred dust 
compositions of SN~2008S and NGC~300-OT, Prieto et al. also conclude 
that most of the dust around NGC~300-OT survived its outburst.

The progenitor of SN\,2008S is estimated to have had an initial mass in 
the range 6--10\,M$_\odot$ (Prieto et al. 2008; Botticella et al. 2009). 
Thompson et al. (2008) suggested that a dust-enshrouded phase may occur 
during the lives of a large fraction of all massive stars, which may 
indicate that this type of transient could be a significant additional 
source of dust in the early universe. Thompson et al. estimated that the 
dust-enshrouded phase of SN~2008S-type objects could last for up to 
2$\times10^4$ years, in which case our derived dust mass-loss rate of 
5.2$\times10^{-7}$~M$_{\odot}$~yr$^{-1}$ implies that up to 
0.01~M$_{\odot}$ of dust could be ejected, along with 1-2~M$_{\odot}$ of 
gas, a significant fraction of the total stellar mass. An ejected 
dust mass of 0.01~M$_{\odot}$ is within a factor of two of the largest 
dust mass derived by any of the recent studies of young CCSNe ejecta 
(SN~2003gd; Sugerman et al. 2006). Up to 1.0~M$_{\odot}$
of cold dust was claimed to be present in the young Galactic supernova 
remnants Cas~A and Kepler by Morgan et al. (2003) and Dunne et al. (2003), 
although their analysis was disputed by Krause et al. (2004). 
Recently Dunne et al. (2009) and Gomez et al. (2009) have returned to 
this issue with new observations, from which they still infer
0.1 - 1.0~M$_{\odot}$ of cold dust in each remnant.

Since the dust around SN~2008S is definitely not silicaceous, and is 
consistent with amorphous carbon, the atoms in the dust will have been 
freshly synthesised, with the outflow enriching the ISM in 
this dust component. Dwek et al. (2007) have estimated that the 
2$\times10^8$~M$_{\odot}$ of dust in the $z = 6.4$ galaxy 
J114816.64+5251 must have been created within a timescale of 400~Myr. 
Given that the lifetime of a 5~M$_{\odot}$ star is only 100~Myr (Maeder \& Meynet 1989) and that 
standard initial mass functions favour lower mass stars over high mass 
stars, then with some initial heavy element seeding from massive CCSNe 
there could be scope for AGB or super-AGB stars with initial masses 
between 6 and 10~M$_{\odot}$, of which the progenitor of SN~2008S appears 
to be an example, to make a contribution to the dust 
enrichment of such galaxies.

We estimate the possible contribution of such stars to the dust content of
J114816.64+5251 as follows: for a Salpeter IMF with limits of 0.1 and
120~M$_{\odot}$, 0.20 per cent of stars have masses between 6 and 10
M$_{\odot}$, while a further 0.29 per cent have masses between 4 and 6
M$_{\odot}$. The star formation history of J114816.64+5251 is uncertain,
with evidence for a current star formation rate of up to 3000
M$_\odot$~yr$^{-1}$ (Dwek et al. 2007).  As they noted, such a high SFR 
almost certainly
could not have been sustained for 400 Myr; $\sim$ 200 M$_\odot$~yr$^{-1}$ is
likely to be a more representative average. This would result in a total
stellar mass of 8$\times$10$^{10}$ M$_\odot$, giving 6.6$\times$10$^8$ 
stars with
masses between 4 and 6 M$_\odot$ and 4.5$\times$10$^8$ stars with masses
between 6 and 10 M$_\odot$.  If each star in both these mass ranges produce
0.01 M$_\odot$ of dust, then their total contribution would be
6.6$\times$10$^6$ and 4.5$\times$10$^6$ M$_\odot$ for the low and high mass
ranges respectively, before dust destruction is accounted for.

A similar calculation can be made for `normal' supernovae; assuming that all
stars with initial masses greater than 8 M$_\odot$ explode as supernovae,
and that each produces 0.01~M$_\odot$ of dust (e.g. Sugerman et al. 2006),
their total contribution before accounting for dust destruction would be
6.0$\times$10$^6$~M$_\odot$.  Thus, the amount of dust produced by
SN\,2008S-type events could be comparable to the contribution of `normal'
supernovae.  For both types of object, these results are sensitive to the
lower bound of the IMF; raising the IMF lower mass limit from
0.1~M$_{\odot}$ to 0.2~M$_\odot$ would approximately double the quantity of
dust produced.

Another parameter that these estimates are sensitive to is the dust 
lifetime against destruction. Some fraction of the dust produced by both 
sources would be destroyed during the galaxy's 400 Myr life. The lifetime 
of dust in this environment, and hence the surviving fraction, is highly 
uncertain; for a typical Milky Way dust lifetime of 5$\times$10$^8$ years 
(Dwek et al. 2007), the surviving fraction after 400\,Myr would be 70\%. 
For star formation bursts as high as 3000 M$_\odot$~yr$^{-1}$, dust 
lifetimes could be as short as 25~Myr (Dwek et al. 2007). However, such 
high star formation rates are unlikely to be sustained for periods longer 
than this, and SN ejecta dust from the massive stars that resulted from 
the burst would seem more at risk from destruction than dust from 
longer-lived high-mass AGB stars, which would be further from the sites of 
active star formation. Even for a dust lifetime of 500~Myr, with 70\% of 
grains surviving over 400~Myr, 4-10~M$_\odot$ AGB stars and SNe would 
respectively account for only 3.9 and 2.1 per cent of the 
2$\times$10$^8$~M$_\odot$ of dust observed in J114816.64+5251. It seems 
difficult for either AGB stars or core collapse SNe to account for the 
masses of dust inferred for such galaxies, if current estimates for 
the galaxy dust masses, and for the dust injection rates by stars, are 
correct.

If SN\,2008S was not a bona fide supernova event, then a star will remain 
at the centre of the dust shell, and could presumably explode later as a 
supernova.  We investigated whether a subsequent supernova could vaporise 
the dust shell, using grids formed from simple geometric expansion of our 
derived dust distribution.  Adopting a representative Type~II supernova 
peak luminosity of 10$^9$L$_\odot$, we find that if a supernova occurred 
in the very near future, then dust within about 7\,500\,AU of the star 
would be vaporised, reducing $\tau_V$ from 0.76 to 0.15.  The surviving 
dust shell would then have a mass of 
3.2$\times$10$^{-3}$~M$_\odot$, still 93\% of the initial value.  If the 
supernova does not occur within the next 500 years, then a 
70\,km\,s$^{-1}$ outflow velocity will have carried the inner edge of the 
shell beyond the vaporisation radius for this luminosity, and the entire 
dust shell would survive the supernova light flash.

\section{Acknowledgments}

We thank the telescope staff for their help in acquiring the 
July 2008 GMOS and Michelle observations (Program ID GN-2008B-Q-44) which 
were obtained at the Gemini Observatory, operated by the 
Association of Universities for Research in Astronomy, Inc., under a 
cooperative agreement with the NSF on behalf of the Gemini partnership: 
the National Science Foundation (United States), the Science and 
Technology Facilities Council (United Kingdom), the National Research 
Council (Canada), CONICYT (Chile), the Australian Research Council 
(Australia), Ministerio da Ciencia e Tecnologia (Brazil) and Ministerio de 
Ciencia, Tecnologia e Innovacion Productiva (Argentina).
This work is based in part on observations made with the Spitzer Space 
Telescope, which is operated by the Jet Propulsion Laboratory, California 
Institute of Technology under a contract with NASA. We thank an
anonymous referee for constructive comments.

\end{document}